\documentstyle[prl,aps,amsfonts,amssymb,twocolumn,epsfig]{revtex}
\newcommand {\be}{\begin{equation}}
\newcommand {\ee}{\end{equation}}
\newcommand {\ba}{\begin{eqnarray}}
\newcommand {\ea}{\end{eqnarray}}

\begin{document}
%\draft

% Uncomment for 2 columns
\twocolumn[\hsize\textwidth\columnwidth\hsize\csname @twocolumnfalse\endcsname
%%%%%%%%%%%%%%%%%%%%%%%%%%%%%%%%%%%%%%%%%%%%%%%%%%%%%%%%%%%%%%%%%%%%%%%%%%%%%%
 
\title{Microscopic chaos from Brownian motion in a 1D anharmonic oscillator chain} 

\author{M.~Romero-Bastida\footnote{Email address: rbm@xanum.uam.mx} and E.~Braun}

\address{Departamento de F\'\i sica, Universidad Aut\'onoma Metropolitana Iztapalapa\\ Apartado Postal 55--534, M\'exico, Distrito Federal, 09340, M\'exico} 

\date {\today}
\maketitle
\begin{abstract}
The problem of relating microscopic chaos to macroscopic behavior in a many-degrees-of-freedom system is numerically investigated by analyzing statistical properties associated to the position and momentum of a heavy impurity embedded in a chain of nearest-neighbor anharmonic Fermi-Pasta-Ulam oscillators. For this model we have found that the behavior of the relaxation time of the momentum autocorrelation function of the impurity is different depending on the dynamical regime (either regular or chaotic) of the lattice.
\end{abstract}
\bigskip
% Uncomment the ``\begin{center},\end{center}'' commands for 2 columns
\begin{center}
{\bf PACS number(s): 05.45.Jn, 05.40.Jc, 05.45.Pq}
\end{center}
% Uncomment for 2 columns
\vskip1.5pc]
% Uncomment for ``preprint'' format
%\newpage
%%%%%%%%%%%%%%%%%%%%%%%%%%%%%%%%%%%%%%%%%%%%%%%%%%%%%%%%%%%%%%%%%%%

\section{Introduction}

A physically relevant and still not satisfactorily solved problem is the question of how the underlying chaotic microscopic dynamics of many-particle systems is related to the observed macroscopic behavior. Assuming that the main features observed at the macroscopic level can be accounted for by the microscopic dynamics, one would expect that some characteristics of the latter, specifically its chaoticity, should be detected, in principle, at the macroscopic level of description.

Recent empirical evidence suggesting microscopic chaos on a molecular scale has been presented in an experiment on the position of a Brownian particle (BP) in a fluid~\cite{gaspard}. The measurements were made at regular time intervals and the experimental time series data was then interpreted using standard techniques of chaotic time series analysis, suggesting a positive lower bound on the Kolmogorov-Sinai entropy, hence, microscopic chaos. However, a similar bound has been obtained with computer experiments on the nonchaotic Eherenfest wind-tree model where a single particle diffuses in a plane due to collisions with randomly placed, fixed, oriented square scatterers~\cite{cohen1}, rendering doubts about the conclusion that microscopic chaos has been experimentaly detected. Further comparisons with the chaotic Lorentz model, which has circular scatterers and has a diffusive behavior that is undistinguishable from the one exhibited by the Eherenfest model, have confirmed that the standard methods of chaotic time series analysis are ill suited to the problem of distinguishing between chaotic and nonchaotic microscopic dynamics, although some alternatives have been proposed in order to overcome this situation~\cite{cohen2}.

The results in Ref.~\cite{gaspard} were attributed to the physical issue of time scales -- the time interval $\Delta t\approx1/60$ s between measurements was vastly greater than the typical collision times $t_c\approx10^{-12}$ s of the BP with the solvent particles in the fluid, which make the diffusive behavior in the experimental data and in the computer experiments virtually identical~\cite{cohen1}. However, there is a time scale in Brownian motion, not mentioned in the previous works, that is characterized by the relaxation time $\tau\approx10^{-8}$ s of the momentum autocorrelation function (MACF) of the BP. Since $\Delta t\gg\tau>t_c$, this relaxation time could, in principle, be used to probe the microscopic dynamics of the system, but so far this possibility has not been explored experimentally. Moreover, the simplicity of the models used in~\cite{cohen1,cohen2} precludes a study in the momentum space from the very beginning, and therefore, a different kind of model is needed to address the problem at hand.

We propose to reconsider a simple Hamiltonian model that has been studied extensively in the past, namely, a one-dimensional chain of harmonic oscillators of unitary mass coupled to a heavy oscillator (impurity). Many-body effects in Brownian motion can be analytically investigated with this model, and it was proved that the impurity of mass $M\gg1$ satisfies a Langevin equation when the thermodynamic limit $N\rightarrow\infty$ is taken and random initial conditions are given to all the oscillators of the system~\cite{homodel}. However, in its original formulation this model is useless to address the proposed problem because of the lack of chaoticity in the lattice, which nevertheless behaves as a heat bath.

In this paper we extend the above-mentioned harmonic model with an anharmonic potential, which makes the resulting Hamiltonian very similar to the Fermi-Pasta-Ulam (FPU) model~\cite{fermi}. With this modification we explore, for this particular model, the possible effect of microscopic chaos in the statistical behavior of the heavy impurity. We note that the problem of an harmonic oscillator coupled linearly to a FPU chain has already been studied~\cite{bianucci}. Diatomic and disordered mass versions of the FPU model have been applied to the problem of energy equipartition~\cite{equi} and heat conduction~\cite{heat}, but not with a single heavy impurity in the context of Brownian motion, as is our case.

\section{The Model}

The Hamiltonian of the model we are considering can be written, in terms of dimensionless variables, as
\be
H\!=\!\!\!\!\!\sum_{i=-N/2}^{N/2} \displaystyle\left[\frac{p_i^2}{2m_i}+\frac{1}{2}(x_{i+1}-x_i )^2 +\frac{1}{4}\beta (x_{i+1}-x_i )^4 \right]~\label{newham},
\ee
\noindent
where $m_i=1$ if $i\not=0$ and $m_0=M$; periodic boundary conditions are assumed ($x_{(N/2)+1}=x_{-N/2}$). The model describes a system of one-dimensional $N$ coupled nonlinear oscillators of unit mass with nearest-neighbor interactions and a central oscillator (impurity) of mass $M$, with displacement $x_0\equiv X$ and momentum $p_0\equiv P$. The quadratic part of Eq.~(\ref{newham}) was the one studied in connection with Brownian motion in Ref.~\cite{homodel}. The value $\beta=0.1$ was used in all the numerical experiments hereafter reported. We call this model the modified FPU (MFPU) model. 

\section{Dynamics in Phase Space}

\begin{figure}[h] 
\epsfig{width=0.95 \linewidth,file=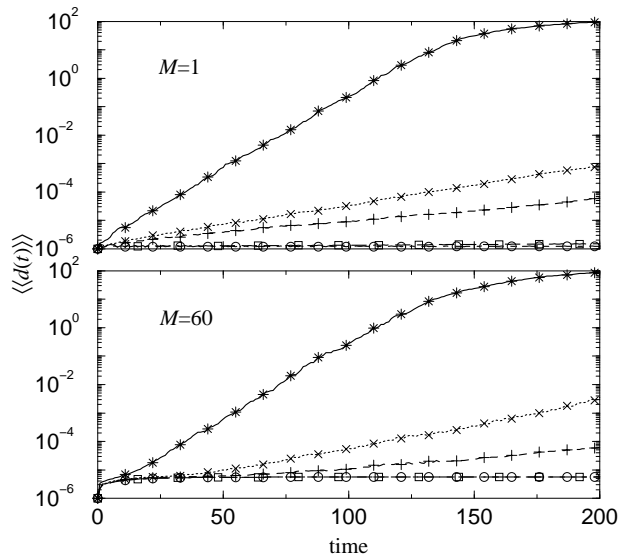}
\caption{Average distance in phase space between pairs of trajectories corresponding to initial conditions such that $d(0)\le10^{-6}$ and $\epsilon=0.01$ (circle), $0.1$ (square), $1$ (plus), $2$ (cross), and $10$ (asterisk) for $M=1$ and $M=60$ with $N=300,000$. Time is measured in natural units.}
\label{fig1}
\end{figure}

It is known that, for high values of the total energy per degree of freedom $\epsilon$, the homogeneous (uniform mass) FPU model [obtained from Eq.~(\ref{newham}) by taking $M=1$] is chaotic whereas, for small $\epsilon$ values, the model behaves as a chain of harmonic oscillators~\cite{pettini}. It is then important to corroborate if these dynamical regimes are in some way affected by the presence of the impurity at lattice site $i=0$. To this end we employ the following distance in phase space
\be
d(t)=\sqrt{\sum_i [\{x^{1}_i(t)-x^{2}_i(t)\}^2+\{p^{1}_i(t)-p^{2}_i(t)\}^2]},
\ee
where the sum runs over all the $N+1$ oscillators of the system, the superscripts $1$ and $2$ refer to two states that at time $t=0$ differ by an infinitesimal quantity $d(0)=10^{-6}$ or less. In our simulations we prepared a system with $N=300,000$ unit mass oscillators in their equilibrium positions. Then we distributed the momenta of the oscillators according to a Maxwell-Boltzmann distribution at a temperature $T$ consistent with the chosen value of $\epsilon$ and let the system evolve in time by solving the equations of motion with an improved leap-frog algorithm. All computations were carried out in double precision. At each temperature an average $\langle\langle d(t)\rangle\rangle$ over ten different realizations was taken to avoid fluctuations that come from a particular choice of the initial conditions. This calculation was performed for different values of $\epsilon$ and $M=1$, $40$, $60$, $80$, and $100$. The results for $M=1$ (shown for comparison) and $60$ are exhibited in Fig.\ \ref{fig1}. As can be seen, an exponential divergence corresponds to values $\epsilon\ge1$ whereas for $\epsilon<1$ the distance is found to be bounded within the same order of magnitude. This happens independently of the value of $M$ for all the cases studied. Thus we can conclude that the qualitative dynamical behavior of the system is only weakly affected by the presence of the heavy impurity. Hence, for small values of $\epsilon$, which correspond to the value range $0.01\le\epsilon<1$ in Fig.\ \ref{fig1}, we find regular (i.e., almost periodic) behavior in phase space whereas for large $\epsilon$ values, which we have studied in the range $1\le\epsilon\le10$, the dynamics in phase space is chaotic~\cite{note1}.

% Comment for ``preprint'' format

\section{Statistical Behavior of the Heavy Impurity}

\begin{figure}[h] 
\epsfig{width=0.95 \linewidth,file=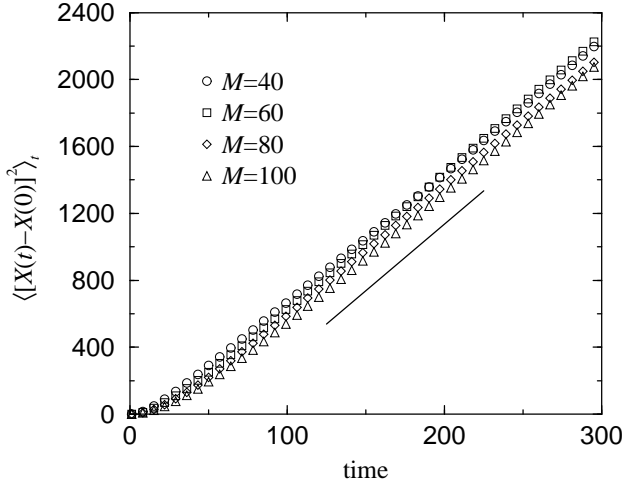}
\caption{Mean-square displacement of the impurity for $\epsilon=10$. The solid line, with slope $2D=7.95$, has been added for reference.}
\label{fig2}
\end{figure}

Thermal equilibrium between the impurity and the FPU chain was attained within the time scale $t=5\times10^5$. Time averages $\langle\ldots\rangle_t$ necessary to obtain the relevant correlation functions were computed afterwards over the characteristic time scale $t=2\times10^5$. The first quantity to be studied was the mean-square displacement. An example is shown in Fig.\ \ref{fig2}; the temperature (i.e., mean kinetic energy per degree of freedom) $T=11.61$ was computed, which is the same value reported in Ref.~\cite{giardina} for the homogeneous FPU model and $\epsilon=10$. For short times ($t<100$) the ballistic behavior of the impurity is evident, whereas, for long times, the diffusive effect of the bath of light oscillators has been established. The noise in the curves reported is negligible for the time interval $100<t<300$, so we can be confident of the values of the ``diffusion coefficients'' $D_{_M}$ obtained from the slope of each $M$ curve in the aforementioned time interval.

% Comment for ``preprint'' format

\begin{figure}[h] 
\epsfig{width=0.85 \linewidth,file=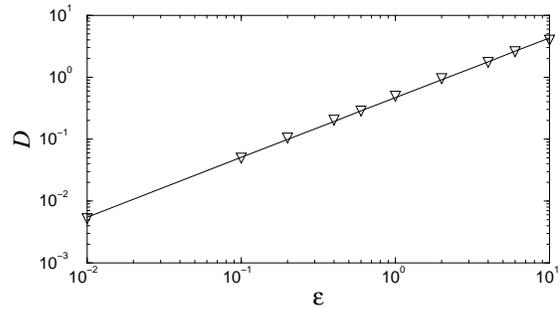}
\caption{Diffusion coefficient vs energy density. The straight line is the power-law fit explained in the text.}
\label{fig3}
\end{figure}

An important issue at this point is to check if the deviations of the individual $D_{_M}$ values from their arithmetic mean $\overline{D}$ are statisticaly significant for any given $\epsilon$; if so, then there is a dependence of the diffusion coefficient on the mass of the heavy oscillator. For a system of coupled harmonic oscillators it is well known that the diffusion coefficient is indeed independent of the mass of the impurity. However, a different situation arises in another type of models. In Ref.~\cite{omerti} it was shown that for a system of noninteracting mass points moving on a line and colliding with a heavy particle there is a mass dependence that increases with the mass of the heavy test particle. For the MFPU model our results show that the deviations $D_{_M}-\overline{D}$ are negligible for small $\epsilon$ values and more pronounced for large $\epsilon$ values. Nevertheless, in both regimes ($\epsilon<1$ and $\epsilon\ge1$) the deviations never exceeded a few percent in all the cases considered, with no systematic drift. Thus we can consider that the diffusion coefficient is given by $D\sim\overline{D}$. For $\epsilon=0.01$ we have obtained the relationship $2D=0.0104\approx T$, which corresponds to the exact result for the harmonic chain~\cite{homodel}. In Fig.\ \ref{fig3} we plot the value of the diffusion coefficient $D$ vs the energy density $\epsilon$. By a least-squares fit we obtain a power-law scaling with respect to $\epsilon$ of the form $D=D_0 \epsilon^m$, where $m=0.964\pm0.008$ and $D_0=0.466\pm0.008$. From the validity of this relationship throughout the entire $\epsilon$ value range studied ($0.01\le\epsilon\le10$), from regular to chaotic, we conclude that, as far as our results allow us to infer, no evidence of the microscopic dynamics can be detected in the behavior of the diffusion coefficient, in agreement with the known results of the Eherenfest and Lorentz models~\cite{cohen1,cohen2}.

% Comment for ``preprint'' format

\begin{figure}[h] 
\epsfig{width=0.85 \linewidth,file=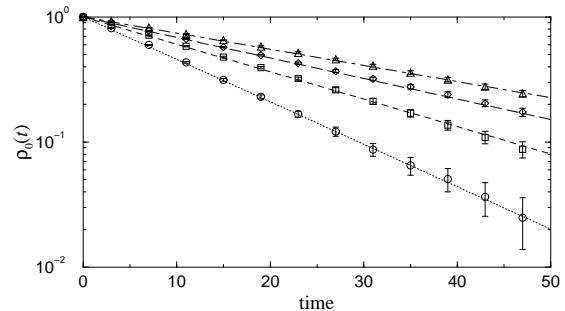}
\caption{Logarithm of the MACF of the impurity for $\epsilon=10$. Symbols have the same meaning as in Fig.\ \ref{fig2}. The straight lines for each value of $M$ correspond to the fit $\exp(-t/\tau)$.}
\label{fig4}
\end{figure}

The MACF $\rho_0(t)\equiv\langle P(t)P(0)\rangle_t/\langle P^2(0)\rangle_t$ of the impurity can also be straightforwardly computed. Figure\ \ref{fig4} shows the time dependence of $\rho_0(t)$ in linear-logarithmic scale for all values of $M$ considered and $\epsilon=10$; other $\epsilon$ values yield similar results. It is evident that, for the relevant time scale ($0\le t\le50$), the MACF exhibits an almost exponential decay, which allows us to obtain the relaxation time $\tau$ from the slope of each curve. The staight lines plotted correspond to the exponential fit $\exp(-t/\tau)$. The observed accuracy of this fit is a clear signature that the impurity indeed performs Brownian motion. The behavior of the MACF is smoother as the mass $M$ increases, so this fact can be considered as another evidence that the more physically relevant results are obtained with larger values of $M$. The magnitude of $\rho_0(t)$ is negligible for $t>50$ in all cases, and consequently its contribution was not considered in the computation of $\tau$.

% Comment for ``preprint'' format

\begin{figure}[h] 
\epsfig{width=0.85 \linewidth,file=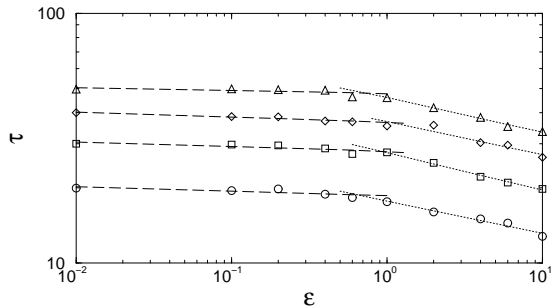}
\caption{Relaxation time vs energy density. The dashed (dotted) lines correspond to the least-squares power-law fit~(\ref{scalinglaw}) for the $\epsilon<1$ ($\epsilon\ge1$) regime and all $M$ values.}
\label{fig5}
\end{figure}

A very interesting result can be obtained if we plot the relaxation time $\tau$ vs the energy density $\epsilon$ for the entire $\epsilon$ value range, from regular to chaotic, and all the considered $M$ values, as shown in Fig.\ \ref{fig5} in log-log scale. We observe that the data points are separated into two different and well-defined regions, depending on the $\epsilon$ value. In all cases the dependence of $\tau$ on the energy density is weak when $\epsilon<1$. On the contrary, when $\epsilon\ge1$, $\tau$ decreases steeply with increasing $\epsilon$. In each of these regimes $\tau$ has a power-law scaling with respect to $\epsilon$ as
\be
\tau_{_M}(\epsilon)=\left\{
\begin{array}{ll}
\tau_{_{0,M}}\epsilon^{\alpha_{_{\! M}}}, & 0.01\le\epsilon<1 \\
\widetilde{\tau}_{_{0,M}}\epsilon^{\widetilde{\alpha}_{_{\! M}}}, & 1\le\epsilon\le10\, .\label{scalinglaw}\\
\end{array}
\right.
\ee
For all the data points corresponding to a fixed $M$ value, $\tau_{_{0,M}}\neq\widetilde{\tau}_{_{0,M}}$ and $\alpha_{_M}\neq\widetilde{\alpha}_{_M}$. This feature contrasts with the behavior of $D$ as seen in Fig.\ \ref{fig3}, where a single scaling law is valid for the entire $\epsilon$ value range. In the $\epsilon<1$ regime the accuracy of the power-law scaling of $\tau_{_M}$ can be tested by noticing that, for $\epsilon=0.01$, $\tau_{_{0,M}}$ must coincide with the exact value $M/2$ of the relaxation time of the MACF for the harmonic chain~\cite{homodel}. This is indeed the case with an accuracy better than $1$\%.

% Comment for ``preprint'' format

\begin{figure}[h] 
\epsfig{width=0.85 \linewidth,file=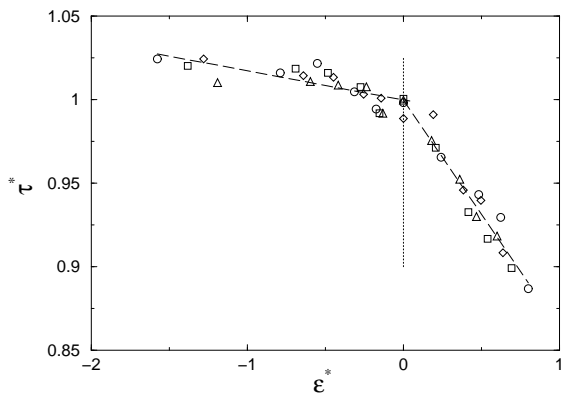}
\caption{Same data as in Fig.\ \ref{fig5}, but in rescaled units $\tau^{\ast}$ and $\epsilon^{\ast}$. Dashed lines are the least-squares fit in each region.}
\label{fig6}
\end{figure}

We notice that the slopes of each of the fits in Fig.\ \ref{fig5} are approximately the same for the $\epsilon<1$ regime. Furthermore the same thing happens for $\epsilon\ge1$, though with other slope value. These facts imply that the scaling exponents in Eq.~(\ref{scalinglaw}) for each regime are approximately the same, independent of the particular $M$ value to which they correspond. It is then reasonable to expect the existence of a mass-independent scaling exponent $\alpha^{\ast}$ ($\widetilde{\alpha}^{\ast}$) for the $\epsilon<1$ ($\epsilon\ge1$) regime. In order to compute $\alpha^{\ast}$ ($\widetilde{\alpha}^{\ast}$) we rescale the data of Fig.\ \ref{fig5} on $\epsilon<1$ ($\epsilon\ge1$) with their corresponding $\tau_{_{0,M}}$ ($\widetilde{\tau}_{_{0,M}}$) value to obtain the new data $\epsilon^{\ast}\equiv\ln\epsilon/\ln\tau_{_{0,M}}$ ($\epsilon^{\ast}\equiv\ln\epsilon/\ln\widetilde{\tau}_{_{0,M}}$) and $\tau^{\ast}\equiv\ln\tau/\ln\tau_{_{0,M}}$ ($\tau^{\ast}\equiv\ln\tau/\ln\widetilde{\tau}_{_{0,M}}$). The result is reported in Fig.\ \ref{fig6}.

% Comment for ``preprint'' format

As can be seen, we have found a common scaling for all the data points in each regime, $\epsilon<1$ and $\epsilon\ge1$. The dashed lines correspond to the best fit of the data in each region, which allow us to calculate the mass-independent scaling exponents $\alpha^{\ast}$ and $\widetilde{\alpha}^{\ast}$. The result is
\be
\tau^{\ast}(\epsilon^{\ast})\sim\left\{
\begin{array}{ll}
(\epsilon^{\ast})^{-0.017\pm0.004}, & -1.57541\le\epsilon^{\ast}<\epsilon_c^{\ast} \\
(\epsilon^{\ast})^{-0.136\pm0.007}, & \epsilon_c^{\ast}\le\epsilon^{\ast}\le0.80121\, .\\
\end{array}
\right.
\ee
$\epsilon_c^{\ast}\approx0$ is the critical value at which the change of scaling occurs and is indicated by the dotted vertical line. The change in behavior when going from low to high $\epsilon$ value is quite pronounced since the scaling exponents differ by one order of magnitude. 

\section{Discussion}

After the presentation of our results we would want to make some remarks. First of all, it could be argued that the diffusion coefficient can be obtained from the MACF by employing the Green-Kubo relation~\cite{boon} that can be written, in terms of the MACF, as~\cite{note2}
\be
D_{_M}=\frac{T}{M}\int_0^{\infty}\rho_0(t)dt\label{gkrelat}.
\ee
This identity implies that $D$ and $\tau$ are not independent variables. More precisely, if we substitute the exponential fit $\rho_0(t)=\exp(-t/\tau)$ in Eq.~(\ref{gkrelat}) we obtain the relation $D_{_M}=(T/M)\tau$, and therefore, the behavior of $\tau$ and $D_{_M}$ as $\epsilon$ increases should be similar. However, we have $T/M\ll1$ for all $M$ values considered, e.g., $T/M=2.5\times10^{-4}$ for $\epsilon=0.01$ and $0.29045$ for $\epsilon=10$ with $M=40$. Thus the behavior of $\epsilon$ vs $\tau$ depicted in Fig.\ \ref{fig5} is lost when $\tau$ is multiplied by the factor $T/M$ to obtain $D_{_M}$. In fact, if we rescale each of the data points displayed in Fig.\ \ref{fig5} with the corresponding factor $T/M$, the result are data sets $\{D_{_M}\}$, one for each $M$ value, spanning the entire $\epsilon$ value range. Furthermore, if we take, for a certain $\epsilon$, the arithmetic mean of all the $D_{_M}$ values, we recover the data points displayed in Fig.\ \ref{fig3}. Thus there is no contradiction between the data in Figs.\ \ref{fig3} and\ \ref{fig5} and the Green-Kubo relation~(\ref{gkrelat}).

To address the problem of the origin of the change in the scaling of $\tau$ as depicted in Fig.\ \ref{fig5} we consider the dynamical behavior of the system as described in Fig.\ \ref{fig1}. We notice that, for the particular value of $M=60$ (the following observations are valid for other $M$ values as well), when $\epsilon<1$ the system is in a regular region of phase space, whereas, for $\epsilon\ge1$, the dynamics is chaotic. This $\epsilon$ value range is approximately the same as that obtained from Fig.\ \ref{fig5} for the change in behavior of $\tau$ as a function of $\epsilon$. From this observation we conclude that the change of scaling in $\tau$ may be attributed to the qualitative change, from regular to chaotic, in the microscopic dynamics of the whole system.

It is important to point out that similar numerical results were obtained from the behavior of the relaxation time $\tau_{_R}$ of the spectral entropy, a function used to study the relaxation to equilibrium of a many-degrees-of-freedom Hamiltonian system starting with far-from-equilibrium initial conditions. For high values of $\epsilon$ (total energy per degree of freedom) $\tau_{_R}\sim\mathrm{const}$, but $\tau_{_R}$ suddenly increases below a certain threshold, $\epsilon_{_R}\sim1$ for the homogeneous FPU model~\cite{pettini}. This result bears some resemblance to the change of scaling in $\tau$ for our model, although the details of behavior as $\epsilon$ decreases are different for both relaxation times $\tau$ and $\tau_{_R}$. Another difference worth stressing is that the relaxation time $\tau_{_R}$ for the homogeneous FPU model is associated to the equilibration process, but the relaxation time $\tau$ reported in Fig.\ \ref{fig5} is the time that characterizes the decay of the MACF of our MFPU model, which is computed in the thermodynamic equilibrium state~\cite{euro}. Moreover, it should be noted that numerical analysis of the dynamics of the homogeneous FPU~\cite{pettini} and other one-dimensional models~\cite{yoshimura} has shown that the change of behavior in $\tau_{_R}$ as $\epsilon$ increases is related to a transition, common to these models, from weak (i.e., almost periodic) to strongly chaotic regimes. It remains an open problem whether the behavior of $\tau$ as reported in Fig.\ \ref{fig5} can be obtained or not when the impurity is coupled to some other system, such as those studied in Ref.~\cite{yoshimura}. The sudden change in the scaling of $\tau$ clearly observed in Fig.\ \ref{fig5} could very well be related to some particular characteristic of the aforementioned transition in the case of the FPU chain. We will address this problem in a future communication.

\section{Conclusions}

We have found, for the particular model studied, that the diffusion coefficient, related to the position of the heavy impurity, is useless as a probe of the microscopic dynamics of the system to which it is coupled. On the contrary, we have presented evidence that the relaxation time of the MACF, related to the momentum of the heavy oscillator, shows a different behavior in going from the regular to the chaotic dynamical regime of the FPU chain as $\epsilon$ increases. To the best of our knowledge this is the first time that such a manifestation of the microscopic dynamics of the FPU model can be observed in the behavior of a physical macroscopic parameter characteristic of Brownian motion. So far we cannot generalize this result to other models. This task will be left for future research.

\acknowledgements

The authors wish to thank M.~A.~Nu\~nez and J.~A.~Gonzalez-Avante for their comments and suggestions. Financial support from Consejo Nacional de Ciencia y Tecnolog\'\i a (CONACyT) M\'exico is also acknowledged.

\vspace*{0.4cm}

\hbox{$^\ast$ Email address: rbm@xanum.uam.mx}


\begin{thebibliography}{99}
\bibitem{note1} We did not study the region $\epsilon>10$ because, as we see from Fig.\ \ref{fig1}, in the interval (1-10) we already have fully developed chaos, in agreement with the results obtained in Ref.~\cite{pettini} for the homogeneous FPU model. Furthermore, there is evidence that the qualitative behavior of the homogeneous FPU model is the same for $\epsilon$ values in the range (1-100)~\cite{giardina}.
\bibitem{note2} The Geen-Kubo relation is usually written as $D=\int_0^{\infty}\langle V(t)V(0)\rangle_t dt$. To obtain Eq.~(\ref{gkrelat}) we have to employ the equipartition principle $M\langle V^2(0)\rangle_t=T$, and so $D_{_M}$.
\end{thebibliography}
\end{document}